\documentclass[11pt]{article}

\usepackage[english]{babel}

\usepackage[letterpaper,top=2cm,bottom=2cm,left=3cm,right=3cm,marginparwidth=1.75cm]{geometry}

\usepackage{cite,enumitem,subfigure}
\usepackage{amssymb}
\usepackage{amsmath,bm,bbm}
\usepackage{siunitx}
\PassOptionsToPackage{hyphens}{url}\usepackage{hyperref}
\usepackage{cleveref}
\usepackage[utf8]{inputenc}
\usepackage[right]{lineno}
\usepackage{csquotes}
\usepackage{booktabs}
\usepackage{longtable}
\usepackage{adjustbox}
\usepackage{array}
\usepackage{url}
\usepackage{appendix}
\usepackage{titlesec}
\usepackage{authblk}
\usepackage{xcolor} 

\usepackage{booktabs,array,multirow}

\def\vh{{\bm{h}}}

\def\vq{{\bm{q}}}
\def\vr{{\bm{r}}}
\def\vs{{\bm{s}}}

\def\vu{{\bm{u}}}
\def\vv{{\bm{v}}}

\def\vy{{\bm{y}}}

\def\mA{{\bm{A}}}

\def\mD{{\bm{D}}}

\def\mF{{\bm{F}}}
\def\mG{{\bm{G}}}
\def\mH{{\bm{H}}}

\def\mM{{\bm{M}}}

\def\mW{{\bm{W}}}

\usepackage{algpseudocode}
\usepackage{algorithm}


\algnewcommand{\IIf}[1]{\State\algorithmicif\ #1\ \algorithmicthen}
\algnewcommand{\EndIIf}{\unskip\ \algorithmicend\ \algorithmicif}

\algnewcommand\algorithmicinput{\textbf{Input:}}
\algnewcommand\algorithmicoutput{\textbf{Output:}}
\algnewcommand\Input{\item[\algorithmicinput]}%
\algnewcommand\Output{\item[\algorithmicoutput]}%

\allowdisplaybreaks

\crefformat{figure}{#2Figure~#1#3}
\Crefformat{figure}{#2Figure~#1#3}
\crefformat{table}{#2Table~#1#3}
\Crefformat{table}{#2Table~#1#3}
\crefformat{section}{#2Section~#1#3}
\Crefformat{section}{#2Section~#1#3}

\author{Shuai Huang, James J. Lah, Jason W. Allen, and Deqiang Qiu\thanks{Shuai Huang is with the Department of Electrical and Computer Engineering, Auburn University, AL 36830 USA (email: shuai.huang@auburn.edu). James J. Lah is with the Department of Neurology, Emory University, Atlanta, GA 30322, USA (jlah@emory.edu). Jason W. Allen is with the Department of Radiology and Imaging Sciences, Indiana University School of Medicine, Indianapolis, IN 46202, USA (email: allenjaw@iu.edu). Deqiang Qiu is with the Department of Radiology and Imaging Sciences, Emory University, GA 30322 USA (email: deqiang.qiu@emory.edu). }}

\title{\bfseries Laplace-Mixture Dipole Inversion for Quantitative Susceptibility Mapping }

\begin{document}
\maketitle

\begin{abstract}
Purpose: To develop an automatic dipole inversion method for quantitative susceptibility mapping (QSM) that preserves fine anatomical structures without the need for manual regularization-parameter tuning.

Theory: The original approximate message passing with parameter estimation (AMP-PE) framework models image gradients with a single Laplace prior, which does not fully capture the heavy-tailed gradient distribution of brain susceptibility maps. This prior mismatch can lead to over-regularization and blocky reconstructions. We address this limitation by modeling the gradients with a two-component Laplace mixture prior.

Methods: We propose a Laplace-Mixture Dipole Inversion (LAMDI) method by incorporating a two-component Laplace mixture prior into the AMP-PE framework with automatic parameter estimation. LAMDI was evaluated on a public in vivo dataset. Its performance was compared with FANSI, MEDI, and AMP-PE with a single-Laplace prior (AMP-PE-L1) under both standard default and reference-tuned settings.

Results: On a public multi-orientation QSM dataset, LAMDI achieved NRMSE and SSIM comparable to AMP-PE-L1 while substantially reducing HFEN, suggesting improved preservation of high-frequency anatomical detail. Under reference-based tuning, FANSI and MEDI achieved the best performance for some metrics, but LAMDI remained competitive without requiring reference maps or manual regularization tuning.

Conclusion: LAMDI provides an effective and automatic parameter-estimation alternative for QSM dipole inversion by combining competitive reconstruction accuracy with improved preservation of fine anatomical detail.

\end{abstract}

\section{Introduction}
\label{sec:introduction}

Quantitative susceptibility mapping (QSM) estimates tissue magnetic susceptibility from gradient-echo phase images \cite{Wang:QSM:2015,Langkammer:QSM:2013,Deistung:QSM_R2Star:2013,Barbosa:QSM_R2Star:2015,Betts:QSM_R2Star:2016,Qiu1085} and provides quantitative contrast related to tissue magnetic composition and microstructure \cite{liu2015quantitative}. It has been widely used to investigate and characterize brain pathologies such as iron-related neurodegeneration (e.g., Parkinson’s and Alzheimer’s disease) \cite{Langkammer:QSM_iron:2012,Schweser:QSM:2012,Li:QSM:2011}, multiple sclerosis lesions \cite{chen2014quantitative,eskreis2015multiple,wisnieff2015quantitative,gillen2021qsm}, intracranial hemorrhage \cite{Zhang:QSM_hemorrhage:2018,Sun:QSM_hemorrhage:2018}, and calcifications \cite{Deistung:QSM_calcification:2013,Chen:QSM_calcification:2014}. In a typical QSM pipeline \cite{schweser2016foundations,chan2021sepia,qsm2024recommended}, the tissue-induced local field is first derived from unwrapped phase images by removing the background field \cite{Liu:PDF:2011,SCHWESER:SHARP:2011}; magnetic susceptibility is then reconstructed from this local field via dipole inversion. Since the dipole kernel becomes singular near the magic angle, this inversion is inherently ill-posed \cite{Wang:QSM:2015,HAACKE20151}. As a result, noise and preprocessing errors can propagate through the reconstruction step and produce characteristic streaking artifacts in the final susceptibility maps, especially near regions with large susceptibility contrasts such as large veins, calcifications, or hemorrhages \cite{wei2015streaking}.

Prior information about the susceptibility map is often incorporated to suppress streaking artifacts. Total-variation (TV) minimization assumes that image gradients are approximately sparse \cite{RUDIN1992259,Strong_2003,Chambolle2004}, i.e. being mostly near zero in homogeneous regions and large primarily at anatomical boundaries. Since streaking artifacts introduce spurious high-frequency variations and excessive large gradients across the entire image, TV minimization could suppress them using this sparse prior by introducing an $l_1$-norm penalty on the image gradients. However, this assumption is only partially valid for brain QSM due to the intrinsically high dynamic range of susceptibility. In particular, physiologic susceptibility variations yield naturally large gradients in and around venous blood, iron-rich deep gray matter nuclei, myelin-rich white matter, and calcifications. Consequently, uniform TV regularization can oversmooth anatomically meaningful contrast and attenuate important structures. To better preserve anatomy, Liu et al. proposed morphology-enabled dipole inversion (MEDI)  \cite{Liu:MEDI:2012}, which uses the magnitude image to identify structural boundaries and reduces gradient penalization across these edges during the inversion. However, this strategy leads to inaccuracies near boundaries, and its performance depends on the quality and design of the morphology edge mask.

A regularization parameter controls the trade-off between data fidelity and the $l_1$-norm regularization term. When a ground-truth reference is available, this parameter can be selected by minimizing reconstruction error. In practice, however, obtaining a reliable reference susceptibility map for tuning is challenging. Although the ``Calculation Of Susceptibility through Multiple Orientations Sampling'' (COSMOS) approach can provide a near–gold-standard reference \cite{liu2009calculation}, it requires phase data acquired at multiple head orientations, making it impractical for routine in vivo studies; moreover, the prolonged scan time increases susceptibility to motion artifacts. Additionally, changes in acquisition protocols or preprocessing often shift the optimal parameter, necessitating retuning. To mitigate these limitations, a number of automatic parameter-selection strategies have been proposed, including automatic selection strategies for L1/TV-regularized QSM \cite{bilgic2014fast}. The L-curve method, for example, plots the log residual norm versus the log regularization norm (e.g., the $l_1$-norm of image gradients) over a range of parameter values and selects the ``corner'', typically defined as the point of maximum curvature, as a compromise between fit and smoothness \cite{Hansen:l_curve:2000,Milovic:PT_QSM:2021}. However, this method is heuristic: the curve is not always distinctly L-shaped, the corner can be ambiguous, and the selected parameter may deviate from the true optimum.

As an alternative probabilistic approach, Huang et al. jointly estimated the susceptibility map and the regularization parameter by maximizing the data likelihood \cite{huang2023robust}. Specifically, they treated the parameter as an unknown and computed its maximum-a-posteriori (MAP) estimate using approximate message passing (AMP) \cite{Rangan:GAMP:2011,PE_GAMP17}. Salman et al. subsequently evaluated the sensitivity of QSM methods for clinical studies of deep gray matter and reported that AMP-PE, HEIDI \cite{schweser2012quantitative}, and LSQR \cite{schweser2012quantitative} achieved the highest overall sensitivity \cite{salman2025sensitivity}. Notably, HEIDI and LSQR require manual parameter tuning, whereas AMP-PE can estimate the parameter automatically. From a probabilistic perspective, AMP-PE models the prior distribution of image gradients using a Laplace distribution. However, because brain susceptibility maps exhibit a high dynamic range, the empirical image gradient distribution can be more heavy-tailed (long-tailed) than Laplacian, leading to a mismatch between the assumed and true priors. This model mismatch can bias the estimated regularization parameter upward, resulting in overly strong regularization that causes susceptibility maps to appear pixelated when zoomed in \cite{salman2025sensitivity}.

In this paper, we propose modeling image gradients with a two-component Laplace mixture prior to better capture their heavy-tailed behavior. Specifically, one component accounts for large-magnitude gradients associated with sharp susceptibility transitions, while the other models small gradients in more homogeneous regions. To incorporate this model into the AMP-PE framework \cite{PE_GAMP17}, we reformulate the parameter-estimation problem using an Expectation–Maximization (EM) procedure \cite{dempster1977maximum}, enabling efficient and straightforward updates of the parameters. Experiments show that the proposed ``Laplace-Mixture Dipole Inversion (LAMDI)'' approach better preserves high-frequency anatomical structures and reduces the blocky or pixelated artifacts observed in the original AMP-PE method with a single-Laplace prior (AMP-PE-L1) \cite{huang2023robust}. We further compare LAMDI with representative dipole inversion methods, including FANSI \cite{Milovic:FANSI:2018,Milovic:PT_QSM:2021} and MEDI \cite{NonlinearMEDI:Liu:2013}, under both standard and reference-tuned settings, and demonstrate that LAMDI provides competitive reconstruction quality without requiring manual parameter tuning.

\section{Theory}

Let $\boldsymbol{\phi}_e$ denote the local field-induced phase\footnote{We assume the initial phase offset $\boldsymbol{\phi}_0$ has been estimated and removed from $\boldsymbol{\phi}_e$.} at the $e$-th echo time $t_e$ in radians, where $e=1,\cdots,E$. It is related to the magnetic susceptibility $\boldsymbol\chi$ by
\begin{align}
\begin{split}
\boldsymbol\phi_e &= 2\pi\gamma\cdot t_e\cdot B_0\cdot\mF^*\mD\mF\boldsymbol\chi\\
&=\mA_e\boldsymbol\chi\,,
\end{split}
\end{align}
where $\gamma$ is the gyromagnetic ratio, $B_0$ is the main magnetic field strength, $\mF$ denotes the Fourier transform operator, $\mD$ is the dipole kernel (singular along the magic angle), and $\mA_e=2\pi\gamma\cdot t_e\cdot B_0\cdot\mF^*\mD\mF$ represents the resulting linear operator.

\begin{figure}[tbp!]
    \centering
    \includegraphics[width=0.7\columnwidth]{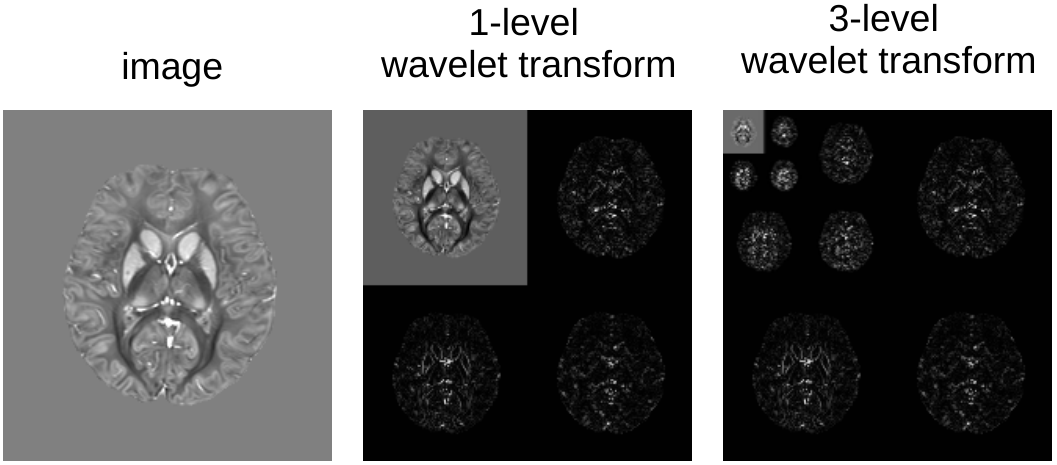}
    \caption{The db1 wavelet transform is applied to an image to produce sparse wavelet coefficients at levels 1 and 3.}
    \label{fig:wavelet_transform}
\end{figure}

\begin{figure}[tbp!]
    \centering
    \includegraphics[width=0.47\columnwidth]{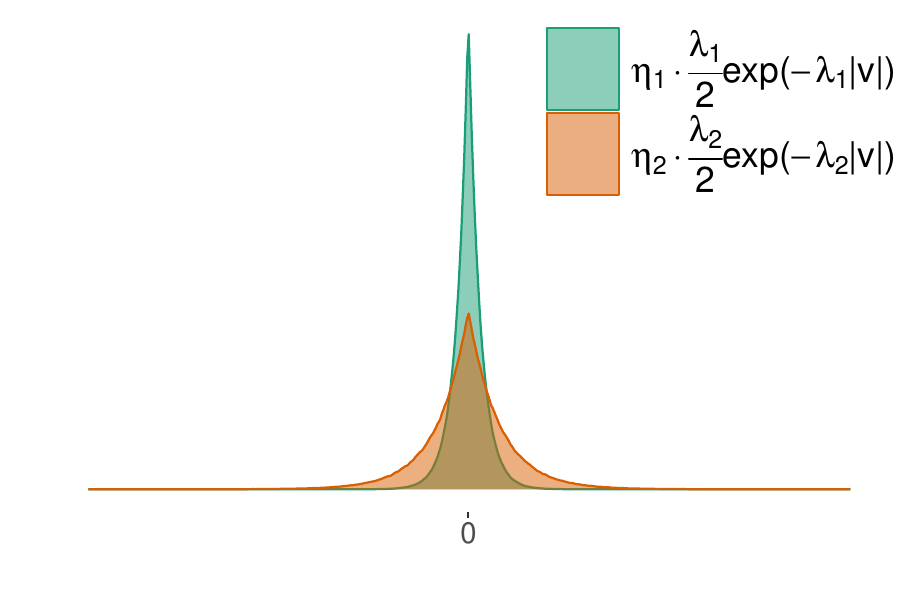}
    \caption{A two-component Laplace mixture distribution is used as the signal prior of image wavelet coefficients.}
    \label{fig:lap_mix}
\end{figure}

We then apply a Haar wavelet transform (db1) to the susceptibility map to obtain wavelet coefficients $\vv$ \cite{DBWav92}, which capture multiscale image differences (i.e., gradient-like detail coefficients) together with a coarse, downsampled approximation. In particular, the low-pass filter $\vh_{\textnormal{low}}=[1\ 1]$ is applied to the image to generate a low-pass, downsampled representation, whereas the high-pass filter $\vh_{\textnormal{high}}=[1\ -1]$ produces gradient-like detail coefficients. As illustrated in Figure \ref{fig:wavelet_transform}, this decomposition can then be applied recursively to the low-pass, downsampled image to obtain higher-level wavelet coefficients. Here we set the decomposition level to 3. The wavelet transform is linear and can be written in a matrix form as follows:
\begin{align}
\vv = \mH\boldsymbol{\chi}\,,
\end{align}
where $\mH$ is the wavelet transform operator (matrix). Owing to the high dynamic range of $\boldsymbol\chi$, the empirical distribution of $\vv$ is strongly heavy-tailed. To model this behavior, we adopt a two-component Laplace-mixture prior distribution for $\vv$.
\begin{align}
\label{eq:lap_mix}
p\left(v|\{\lambda_s,\eta_s\}_{s=1}^2\right)=\sum_{s=1}^2\eta_s\cdot\frac{\lambda_s}{2}\exp(-\lambda_s|v|)\,,
\end{align}
where $\eta_s$ is the $s$-th mixture weight, and $\lambda_s$ is $s$-th Laplacian parameter. As shown in Figure \ref{fig:lap_mix}, the first Laplace component captures small gradients, whereas the second component models the large-magnitude gradients that constitute the heavy-tailed portion of the empirical distribution.

To enable robust estimation of the susceptibility map $\boldsymbol\chi$, we adopt the nonlinear forward model in \cite{NonlinearMEDI:Liu:2013}, where the complex measurement $\vy_e$ at the $e$-th echo is written as
\begin{align}
\label{eq:nonlinear_model}
\begin{split}
\vy_e&=\mW_e\cdot\exp(i\boldsymbol\phi_e)=\mW_e\cdot\exp(i\mA_e\boldsymbol\chi)+\vu\,,
\end{split}
\end{align}
where $i$ denotes the imaginary unit, $\mW_e$ is a diagonal weighting matrix initialized by the magnitude image $\mM_e$ and $\vu$ is complex additive white Gaussian noise, 
\begin{align}
\vu\sim\mathcal{CN}(\vu|0,\tau)=\frac{1}{\pi\tau}\cdot\exp\left(-\frac{|\vu|^2}{\tau}\right),
\end{align}
where $\tau$ is the noise variance.

In practice, we linearize the nonlinear model in \eqref{eq:nonlinear_model} iteratively using a first-order Taylor approximation (including the zeroth-order term) to obtain a computationally tractable estimation problem. The resulting linearized measurement model is detailed in Appendix~\ref{app:sec:linearization}. In the $l$-th iteration, we then have:
\begin{align}
\vy_e^{(l)}=\mA_e^{(l)}\boldsymbol\chi + \vu\,,
\end{align}
where $\vy_e^{(l)}$ and $\mA_e^{(l)}$ are the linearized measurement and operator respectively. We jointly estimate the susceptibility map $\boldsymbol\chi$ and the distribution parameters 
$\eta_s,\lambda_s,\tau$ within the approximate message passing with parameter estimation (AMP-PE) framework \cite{PE_GAMP17}. 



In AMP-PE, the unknown signal is represented by the wavelet coefficient vector 
$\vv$ of the susceptibility map $\boldsymbol{\chi}$. Since $\vv = \mH \boldsymbol{\chi}$, the susceptibility 
map can be written as $\boldsymbol{\chi} = \mH^{-1}\vv$. Therefore, after linearization of the 
nonlinear QSM signal model at outer iteration $l$, the measurement model for the 
$e$-th echo can be expressed in the wavelet domain as
\begin{align}
    \vy_e^{(l)} 
    = \widetilde{\mA}_e^{(l)} \vv + \vu ,
\end{align}
where
\begin{align}
    \widetilde{\mA}_e^{(l)} = \mA_e^{(l)} \mH^{-1}.
\end{align}
For notational simplicity, we drop the echo index $e$ and the outer-iteration 
index $(l)$ in the AMP updates and denote the corresponding wavelet-domain 
measurement operator by $\widetilde{\mA}$.

In practice, damping was applied to improve the stability of the AMP iterations 
for the ill-conditioned and highly structured dipole inversion operator
\cite{Rangan:DampingCvg:2014,Vila:DampingMR:2015}. Let $\vv^{(t)}$ denote the 
damped estimate of the wavelet coefficients at AMP iteration $t$, and let 
$\widehat{\vv}^{(t+1)}$ denote the undamped estimate produced by the AMP nonlinear 
input update at iteration $t+1$. The damped update is then given by
\begin{align}
\label{eq:wavelet_coeff_damping}
    \vv^{(t+1)}
    = \vv^{(t)} + \alpha
    \left(\widehat{\vv}^{(t+1)}-\vv^{(t)}\right),
\end{align}
where $\alpha\in (0,1]$ is the damping rate for the wavelet coefficient update. 
The choice $\alpha = 1$ corresponds to the standard undamped AMP update, whereas 
smaller values yield more conservative updates. For the QSM dipole inversion 
experiments, we used a small damping rate, $\alpha \leq 0.01$, which was found 
empirically to improve convergence stability. The detailed algorithm is provided in Appendix~\ref{app:sec:amp_pe}.

\subsection{Laplace-mixture dipole inversion}

The prior distribution of the wavelet coefficients $v$ is modeled as a two-component Laplace mixture $p\left(v|\{\lambda_s,\eta_s\}_{s=1}^2\right)$ in \eqref{eq:lap_mix}. Within AMP-PE, the variable node associated with $v$ receives an (approximately) Gaussian message $\mathcal{N}(v|r,\tau_r)$ from the output channel, with mean $r$ and variance $\tau_r$. Combining this message with the Laplace-mixture prior yields the (unnormalized) marginal posterior
\begin{align}
    p(v|\vy)\propto \mathcal{N}(v|r,\tau_r)\cdot p\left(v|\{\lambda_s,\eta_s\}_{s=1}^2\right).
\end{align}

We estimate $v$ by computing its maximum a posteriori (MAP) solution, i.e., by maximizing $\log p(v|\vy)$. Ignoring constant terms, this is equivalent to maximizing
\begin{align}
-\frac{1}{2\tau_r}(v-r)^2+\log\sum_{s=1}^2 \eta_s,\frac{\lambda_s}{2}\exp(-\lambda_s|v|).
\end{align}
The second log-sum term is non-convex and difficult to optimize directly. To obtain a tractable update, we linearize this term with respect to $v$ at $v=r$, which leads to a reweighted-$l_1$ minimization problem:
\begin{align}
\hat{v}=\arg\min_v \frac{1}{2\tau_r}(v-r)^2+w(r)\cdot|v|,
\end{align}
where the weight $w$ is given by
\begin{align}
w(r)=\frac{\sum_{s=1}^2 \eta_s \lambda_s^2 \exp(-\lambda_s|r|)}
{\sum_{s=1}^2 \eta_s \lambda_s \exp(-\lambda_s|r|)}.
\end{align}
The resulting MAP estimate of $v$ is given by the soft-thresholding solution
\begin{align}
\label{eq:est_wave_coef}
\hat{v}=\textnormal{sign}(r)\cdot \max\left(|r|-\tau_r w(r),\ 0\right).
\end{align}

\section{Methods}

We compared the proposed LAMDI method with three representative dipole inversion approaches: the established MEDI \cite{NonlinearMEDI:Liu:2013} and FANSI \cite{Milovic:FANSI:2018,Milovic:PT_QSM:2021}, and our previous AMP-PE framework with a single-Laplace prior (denoted AMP-PE-L1) \cite{huang2023robust}.

\begin{enumerate}
    \item \textbf{FANSI} solves the following nonlinear dipole inversion problem with a total variation (TV) regularizer:
    \begin{align}
        \label{eq:fansi}
        \textnormal{FANSI:}\quad
        \min_{\boldsymbol\chi}\ \left\|\mM_e\left(\exp\left(i\mA_e\boldsymbol\chi\right)-\exp\left(i\boldsymbol\phi_e\right)\right)\right\|_2^2+\zeta\|\mG\boldsymbol\chi\|_1,
    \end{align}
    where $\mM_e$ is a magnitude-based weighting matrix for the nonlinear phase data-fidelity term, $\mG(\cdot)$ denotes the image-gradient operator, and $\zeta$ controls the trade-off between data fidelity and TV regularization. Following the recommendations in the FANSI toolbox \cite{Milovic:PT_QSM:2021}, we selected $\zeta$ using the L-curve method over the parameter set
    \[
    \left\{10^{\left(-1.5-(i+20)\times 0.1\right)} \,\middle|\, i=1,\dots,25 \right\}.
    \]
    The curvature of the L-curve was evaluated for all candidate values, and the parameter corresponding to the point of maximum curvature was chosen as the optimal $\zeta$. The maximum number of iterations was set to 500, with a convergence tolerance of $10^{-3}$.

    \item \textbf{MEDI} solves a related nonlinear dipole inversion problem of the form
    \begin{align}
        \label{eq:medi_obj}
        \textnormal{MEDI:}\quad
        \min_{\boldsymbol\chi}\ \rho\left\|\mW_e\left(\exp\left(i\mA_e\boldsymbol\chi\right)-\exp\left(i\boldsymbol\phi_e\right)\right)\right\|_2^2+\|\mathcal{M}_g\mG\boldsymbol\chi\|_1,
    \end{align}
    where $\mathcal{M}_g$ is a binary morphology mask that selects non-edge gradients for regularization, and $\rho$ is the regularization parameter. Because the weighting matrix $\mW_e$ is iteratively updated according to the residual error, it changes with $\rho$, thereby introducing additional variability into the data-fidelity term and altering the shape of the L-curve. For this reason, we did not use the L-curve method for MEDI in the standard-parameter comparison. We therefore used the default value $\rho=1000$ recommended in the MEDI toolbox \cite{NonlinearMEDI:Liu:2013}.

    \item \textbf{AMP-PE-L1} assumes that the wavelet coefficients in $\vv$ follow a single Laplace distribution:
    \begin{align}
        p(v|\lambda)=\frac{\lambda}{2}\exp(-\lambda|v|),
    \end{align}
    where the distribution parameter $\lambda$ is automatically estimated by maximizing the posterior distribution $p(\lambda|\vy)$. However, a single Laplace distribution does not adequately capture the empirically heavy-tailed distribution of the wavelet coefficients. As a result, $\lambda$ tends to be overestimated, leading to over-regularization and susceptibility maps with a blocky or pixelated appearance. 
\end{enumerate}

\subsection{Multi-orientation GRE QSM dataset}

We used the publicly available multi-orientation gradient-echo MRI dataset released by Shi \textit{et al.} to benchmark the aforementioned dipole inversion approaches \cite{shi2022towards}. The dataset contains 144 scans from 8 healthy subjects and includes raw phase/magnitude data, averaged local field maps, COSMOS reconstructions, six symmetric susceptibility tensor components \cite{li2017susceptibility}, and deep gray matter parcellation maps. The scans from multiple head orientations enable the reconstruction of high-quality susceptibility references via COSMOS for evaluating single-orientation dipole inversion algorithms.

All healthy subjects were scanned on a 3T Prisma scanner (Siemens Healthineers, Erlangen, Germany) using a multi-echo 3D GRE sequence with $1 \times 1 \times 1$ mm$^3$ spatial resolution, $TR=44$ ms, $TE=7.7/13.4/18.8/25.3/31.7/38.2$ ms, flip angle $20^\circ$, bandwidth 190 Hz/pixel, field of view $210 \times 224 \times 160$ mm$^3$, matrix size $210 \times 224 \times 160$, and GRAPPA factor 2. In the original preprocessing pipeline based on the STISuite toolbox (\href{https://chunleiliulab.github.io/software/}{https://chunleiliulab.github.io/software/}), brain mask was generated using FSL BET \cite{Smith:BET:2002}, phase unwrapping was performed using a Laplacian-based method \cite{schofield2003fast}, background field removal was performed with V-SHARP \cite{wu2012whole}. The averaged local field maps from different orientations were aligned using FSL FLIRT before reconstructing COSMOS and an STI-based susceptibility references. Three evaluation metrics are computed with respect to the COSMOS-based reference susceptibility maps: normalized root mean squared error (NRMSE), high-frequency error norm (HFEN) and structural similarity index measure (SSIM). These metrics were chosen to provide global-error, high-frequency-error, and structural-similarity assessments, consistent with QSM reconstruction-challenge evaluations \cite{langkammer2018quantitative}.

\begin{itemize}
\item We first used the echo-averaged local field maps (in Hz) provided by the dataset to reconstruct susceptibility maps with all four methods. This experiment was intended to establish a common baseline for comparing the different dipole inversion approaches under the same input setting. For consistency with the nonlinear dipole inversion formulations used by these methods, the echo-averaged local field maps were converted to the phase domain (in radians) by assuming a single echo time of $TE=8$ ms\footnote{We used TE = 8 ms as a representative effective echo time for converting the echo-averaged local field maps to phase, consistent with the range commonly used in the tested nonlinear inversion toolboxes.}. The resulting phase maps were then used as the inputs for FANSI, MEDI, AMP-PE-L1, and the proposed LAMDI.

\item The above comparison reflects a practical setting in which FANSI and MEDI are used with standard default parameter-selection strategies. To further compare the proposed method against reference-tuned classical baselines, we conducted an additional experiment in which the regularization parameters of FANSI and MEDI were tuned on a separate training set of 5 subjects (S1--S5) using COSMOS-based reference susceptibility maps. The four dipole inversion methods were then compared on a held-out test set of 3 subjects (S6--S8). In contrast, AMP-PE-L1 and LAMDI were applied directly to the test subjects without parameter tuning, since both methods estimate their parameters automatically. This setting gives FANSI and MEDI access to information that is not used by the automatic AMP-based methods. It allows us to assess the competitiveness of the proposed automatic approaches against reference-tuned baselines.

In preliminary experiments, we found that tuning FANSI using NRMSE alone was not sufficient. Although this strategy reduced the global reconstruction error, it often selected an overly large regularization parameter, leading to oversmoothed susceptibility maps with noticeable loss of fine anatomical detail. By comparison, NRMSE-based tuning worked well for MEDI. To adopt a consistent tuning strategy for both methods, we therefore selected the parameters by minimizing the sum of NRMSE and HFEN on the training set. This yielded an optimal parameter of \(\zeta=10^{-4}\) for FANSI and \(\rho=500\) for MEDI. 

\end{itemize}

\clearpage

\section{Results}

\subsection{Comparison under standard parameter-selection settings}
\begin{figure}[tbp!]
    \centering
    \includegraphics[width=\columnwidth]{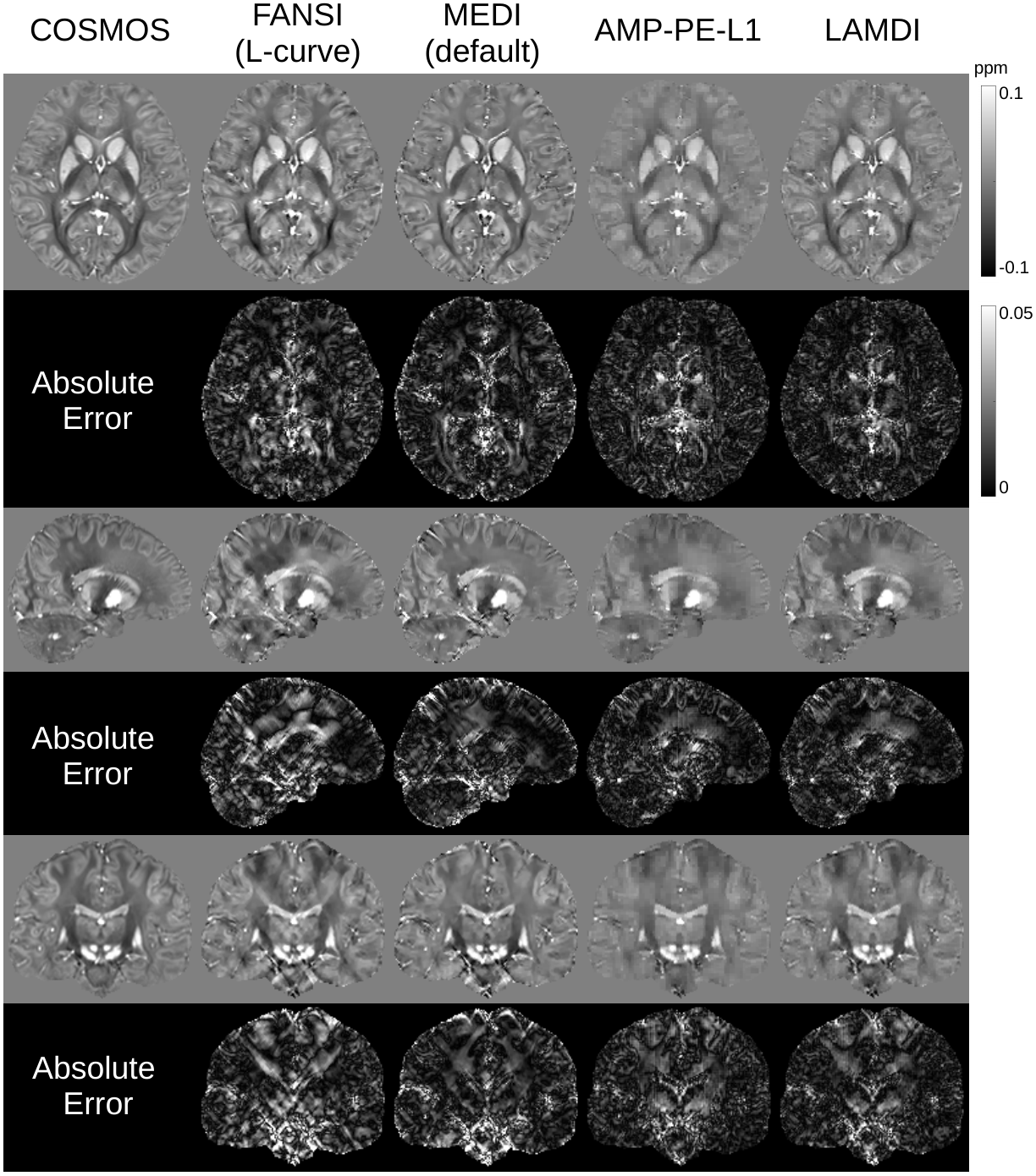}
    \caption{Susceptibility maps reconstructed using COSMOS, FANSI with L-curve–based parameter selection, MEDI with the default parameter setting, AMP-PE-L1 and the proposed LAMDI with automatic parameter estimation. The corresponding absolute error maps were computed with respect to the COSMOS reference susceptibility maps.}
    \label{fig:compare_approaches}
\end{figure}

\begin{figure*}[tbp]
\centering
\subfigure[]{
\label{fig:barplot_standard_setting}
\includegraphics[width=0.47\textwidth]{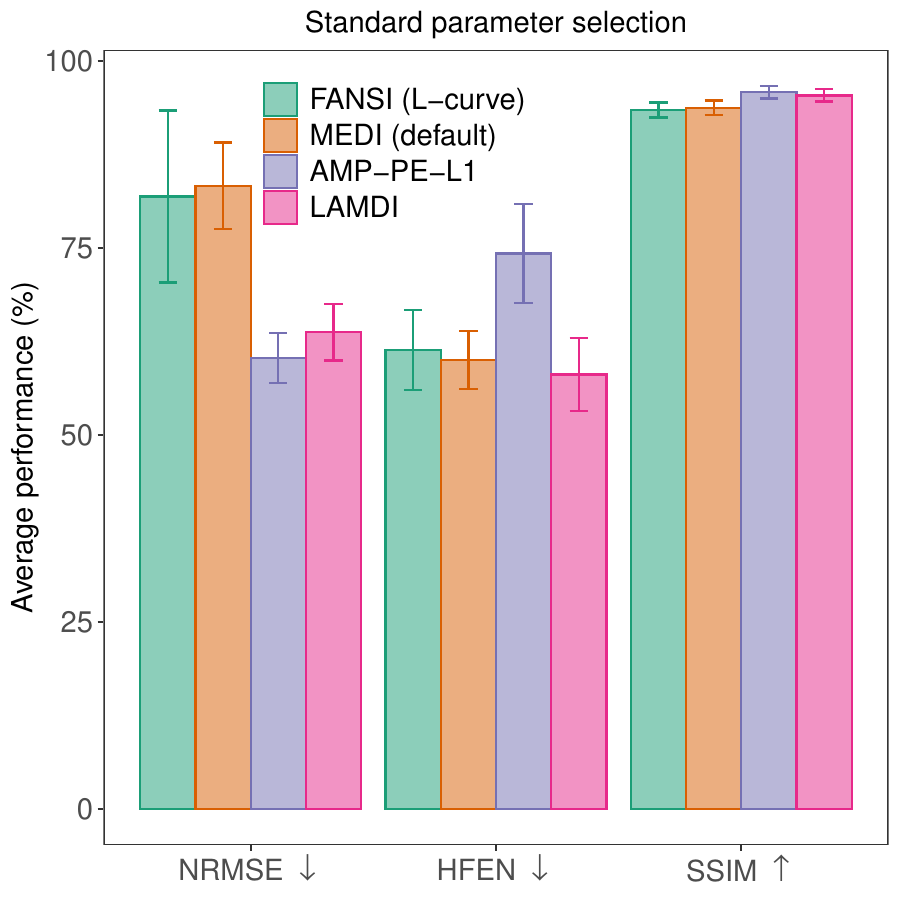}}
\subfigure[]{
\label{fig:barplot_tuned_setting}
\includegraphics[width=0.47\textwidth]{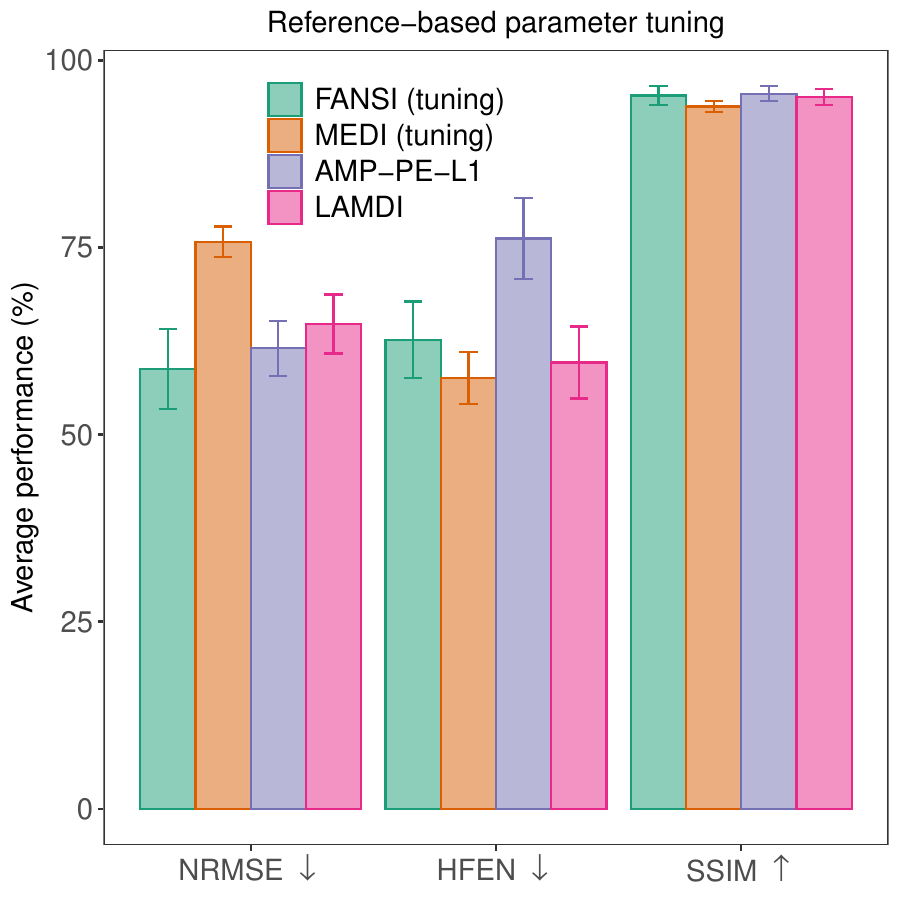}}
\caption{Bar plots of the average quantitative performance of the four dipole inversion methods under the standard and reference-tuned settings. ``$\downarrow$'' indicates lower values are better, ``$\uparrow$'' indicates higher values are better. Error bars denote standard deviations. (a) In the standard setting, FANSI used L-curve–based parameter selection and MEDI used the default parameter setting. (b) In the reference-tuned setting, FANSI and MEDI were tuned on the training subjects (S1--S5) using COSMOS-based reference susceptibility maps. AMP-PE-L1 and LAMDI estimated their parameters automatically in both settings.}
\label{fig:barplot_standard_and_tuned_settings}
\end{figure*}

Figure~\ref{fig:compare_approaches} shows the reconstructed susceptibility maps from Subject 6 using the four dipole inversion methods. The absolute error maps were calculated with respect to the reference maps produced by COSMOS. Table~\ref{tab:main_results} and Figure~\ref{fig:barplot_standard_setting} summarize the quantitative comparison on the eight healthy subjects in the public multi-orientation QSM dataset. Overall, the two AMP-based methods achieved lower NRMSE and higher SSIM than FANSI and MEDI under the standard parameter-selection settings. In particular, AMP-PE-L1 achieved the lowest average NRMSE (\(60.30\%\)) and the highest average SSIM (\(95.83\%\)), while LAMDI produced comparable values, with an average NRMSE of \(63.74\%\) and an average SSIM of \(95.42\%\). By contrast, FANSI and MEDI yielded much larger NRMSE values (\(81.89\%\) and \(83.32\%\), respectively) and lower SSIM values (\(93.48\%\) and \(93.75\%\), respectively). These results indicate that both AMP-based methods provide better global reconstruction fidelity relative to the conventional TV-based and morphology-constrained baselines.

A different trend was observed for HFEN, which is more sensitive to high-frequency image content and local structural detail. FANSI, MEDI, and the proposed LAMDI achieved broadly comparable HFEN values, with LAMDI attaining the lowest average HFEN (\(58.08\%\)), slightly outperforming both FANSI (\(61.38\%\)) and MEDI (\(60.07\%\)). It is particularly noteworthy that LAMDI yielded a markedly lower HFEN than AMP-PE-L1 (\(74.28\%\)). Thus, although AMP-PE-L1 and LAMDI performed comparably in terms of NRMSE and SSIM, LAMDI preserved fine anatomical structures more effectively. This finding is important because it suggests that LAMDI maintains the global reconstruction accuracy of AMP-PE-L1 while addressing its known tendency to lose high-frequency detail.

Post-hoc paired Wilcoxon signed-rank tests between the best-performing method and the remaining methods were performed for each evaluation metric, and the results are shown in Table S1 in the Supporting Information. AMP-PE-L1 achieved significantly lower NRMSE than FANSI, MEDI, and LAMDI, with a p-value $p=0.0078$ for all three comparisons. For HFEN, LAMDI achieved the lowest average value, but the differences from FANSI and MEDI were not statistically significant ($p=0.1953$ and $p=0.0547$, respectively), while the difference from AMP-PE-L1 was significant ($p=0.0078$). For SSIM, AMP-PE-L1 was significantly higher than FANSI, MEDI, and LAMDI, with $p=0.0078$ for all three comparisons.

\begin{table}[tbp]
\caption{Quantitative comparison of the four dipole inversion methods on the eight subjects (S1--S8). ``$\downarrow$'' indicates lower values are better, ``$\uparrow$'' indicates higher values are better. FANSI used L-curve–based parameter selection, MEDI used the default parameter setting, and AMP-PE-L1 and LAMDI estimated their parameters automatically. Best results for each metric are shown in bold.}
\vspace{1em}
\label{tab:main_results}
\centering
\resizebox{\columnwidth}{!}{
\begin{tabular}{llccccccccc}
\toprule
Metric($\%$) &Method &Ave. &S1 &S2 &S3 &S4 &S5 &S6 &S7 &S8\\ \cmidrule(lr){1-2} \cmidrule(lr){3-3} \cmidrule(lr){4-11}
&FANSI (L-curve) &81.89$\pm$11.51 &66.8 &73.5 &80.8 &83.5 &106.8 &82.4 &80.0 &81.3 \\
&MEDI (default) &83.32$\pm$5.80 &72.2 &81.4 &88.7 &82.2 &91.3 &80.4 &84.9 &85.3 \\
&AMP-PE-L1 &\textbf{60.30$\pm$3.34} &\textbf{56.2} &\textbf{58.9} &\textbf{65.0} &\textbf{57.8} &\textbf{59.9} &\textbf{58.3} &\textbf{65.5} &\textbf{60.7} \\
\multirow{-4}{*}{NRMSE $\downarrow$} &LAMDI &63.74$\pm$3.77 &56.6 &64.6 &67.2 &63.0 &64.2 &61.6 &69.2 &63.5   \\
\cmidrule(lr){1-2} \cmidrule(lr){3-3} \cmidrule(lr){4-11}
&FANSI (L-curve) &61.38$\pm$5.35 &52.1 &\textbf{57.1} &\textbf{63.1} &62.0 &70.6 &60.0 &\textbf{63.7} &62.5 \\
&MEDI (default) &60.07$\pm$3.88 &52.6 &59.4 &63.6 &59.1 &62.8 &57.3 &64.4 &61.3 \\
&AMP-PE-L1 &74.28$\pm$6.60 &69.3 &71.6 &86.1 &66.6 &72.2 &71.2 &81.8 &75.6 \\
\multirow{-4}{*}{HFEN $\downarrow$} &LAMDI &\textbf{58.08$\pm$4.87} &\textbf{51.3} &57.2 &65.5 &\textbf{55.9} &\textbf{55.9} &\textbf{55.2} &64.7 &\textbf{59.0}  \\
\cmidrule(lr){1-2} \cmidrule(lr){3-3} \cmidrule(lr){4-11}
&FANSI (L-curve) &93.48$\pm$1.00 &94.9 &94.8 &92.5 &93.6 &92.5 &93.7 &92.3 &93.6 \\
&MEDI (default) &93.75$\pm$0.98 &94.8 &94.6 &92.1 &94.2 &94.1 &94.2 &92.5 &93.7 \\
&AMP-PE-L1 &\textbf{95.83$\pm$0.81} &\textbf{96.4} &\textbf{96.4} &\textbf{94.7} &\textbf{96.1} &\textbf{96.4} &\textbf{96.2} &\textbf{94.4} &\textbf{96.0}  \\
\multirow{-4}{*}{SSIM $\uparrow$} &LAMDI &95.42$\pm$0.81 &96.2 &95.9 &94.4 &95.6 &96.0 &95.8 &93.9 &95.7  \\
\bottomrule
\end{tabular}
}
\end{table}

\begin{figure}[tbp!]
    \centering
    \includegraphics[width=\columnwidth]{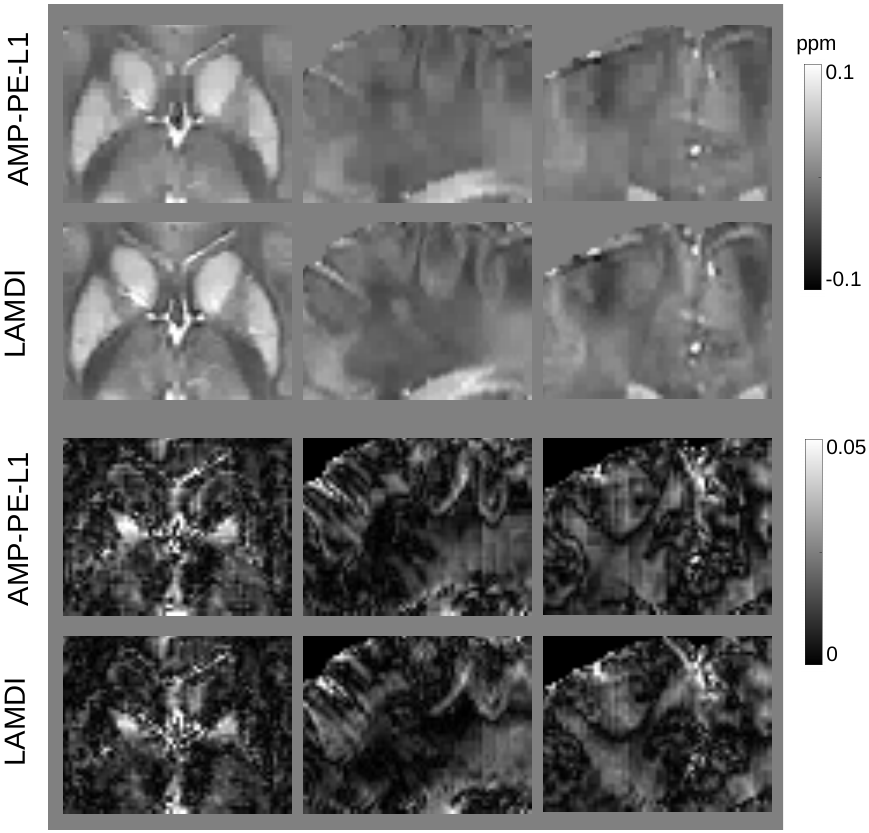}
    \caption{Comparison of susceptibility maps (and absolute error maps) reconstructed by AMP-PE-L1 and LAMDI in representative regions across the axial, sagittal and coronal views. In the selected zoomed regions, AMP-PE-L1 produces a block-like or pixelated appearance, whereas LAMDI reduces the block-like artifact and preserves more visible fine detail.}
    \label{fig:compare_amp_pe_lamdi}
\end{figure}

Figure~\ref{fig:compare_amp_pe_lamdi} provides a visual comparison between the susceptibility maps reconstructed by AMP-PE-L1 and the proposed LAMDI method. Overall, both methods recover the major anatomical structures consistently, but noticeable differences can be observed in the zoomed regions. In the AMP-PE-L1 reconstruction, these regions exhibit a more blocky or pixelated appearance, with local intensity transitions that appear more abrupt and fine structural patterns that are less well preserved. By contrast, the LAMDI reconstruction appears visually smoother while still maintaining sharper anatomical boundaries and more coherent tissue contrast. These differences are visible in the cortical and deep-brain regions across the axial, sagittal, and coronal views. The visual comparison is consistent with the quantitative HFEN results, suggesting that the Laplace-mixture prior in LAMDI better preserves high-frequency anatomical detail while alleviating the blocky artifacts observed in AMP-PE-L1.

\begin{table}[tbp]
\caption{Quantitative comparison of reference-tuned FANSI, MEDI and automatic AMP-based methods on the held-out test subjects (S6--S8). ``$\downarrow$'' indicates lower values are better, ``$\uparrow$'' indicates higher values are better. FANSI and MEDI were tuned on the training subjects (S1--S5) using COSMOS-based reference susceptibility maps, whereas AMP-PE-L1 and LAMDI estimated their parameters automatically without training. Best results for each metric are shown in bold.}
\vspace{1em}
\label{tab:comparison_tuned}
\centering
\begin{tabular}{llccccccccc}
\toprule
Metric($\%$) &Method &Ave. &S6 &S7 &S8\\ \cmidrule(lr){1-2} \cmidrule(lr){3-3} \cmidrule(lr){4-6}
&FANSI (tuning) &\textbf{58.77$\pm$5.31}  &\textbf{55.0} &\textbf{64.9} &\textbf{56.4}\\
&MEDI (tuning) &75.78$\pm$2.02 &73.6 &76.3 &77.5 \\
&AMP-PE-L1 &61.52$\pm$3.66 &58.3 &65.5 &60.7 \\
\multirow{-4}{*}{NRMSE $\downarrow$} &LAMDI &64.78$\pm$3.94 &61.6 &69.2 &63.5  \\
\cmidrule(lr){1-2} \cmidrule(lr){3-3} \cmidrule(lr){4-6}
&FANSI (tuning) &62.66$\pm$5.11 &56.8 &65.5 &65.7 \\
&MEDI (tuning) &\textbf{57.54$\pm$3.47} &\textbf{54.0} &\textbf{60.9} &\textbf{57.7} \\
&AMP-PE-L1 &76.20$\pm$5.37 &71.2 &81.8 &75.6 \\
\multirow{-4}{*}{HFEN $\downarrow$} &LAMDI &59.64$\pm$4.81 &55.2 &64.7 &59.0 \\
\cmidrule(lr){1-2} \cmidrule(lr){3-3} \cmidrule(lr){4-6}
&FANSI (tuning) &95.30$\pm$1.24 &96.0 &93.9 &\textbf{96.0} \\
&MEDI (tuning) &93.84$\pm$0.71 &94.5 &93.1 &94.0  \\
&AMP-PE-L1 &\textbf{95.55$\pm$0.98} &\textbf{96.2} &\textbf{94.4} &\textbf{96.0} \\
\multirow{-4}{*}{SSIM $\uparrow$} &LAMDI &95.13$\pm$1.05 &95.8 &93.9 &95.7  \\
\bottomrule
\end{tabular}
\end{table}

\subsection{Comparison under reference-based parameter tuning}

The quantitative results on the three held-out test subjects are summarized in Figure~\ref{fig:barplot_tuned_setting} and Table~\ref{tab:comparison_tuned}. Under this reference-tuned setting, FANSI achieved the best average NRMSE (\(58.77\%\)) and highly competitive SSIM (\(95.30\%\)), corresponding to the lowest NRMSE among the evaluated methods. MEDI achieved the best average HFEN (\(57.54\%\)), corresponding to the lowest HFEN among the evaluated methods after tuning. These results show that, when COSMOS-based reference maps are available for parameter calibration, classical dipole inversion methods can perform very well when reference-based parameter tuning is available.

The two AMP-based methods exhibited a different performance profile. AMP-PE-L1 achieved the highest average SSIM (\(95.55\%\)) and the second-best NRMSE (\(61.52\%\)), but its HFEN remained worse (\(76.20\%\)), consistent with its tendency to produce blocky or over-regularized reconstructions. LAMDI yielded average NRMSE, HFEN, and SSIM values of \(64.78\%\), \(59.64\%\), and \(95.13\%\), respectively. Although LAMDI did not surpass the tuned FANSI or MEDI baselines in this experiment, it remained close to the best HFEN and SSIM values, while requiring no training set or manual parameter adjustment.

\begin{figure}[tbp!]
    \centering
    \includegraphics[width=\columnwidth]{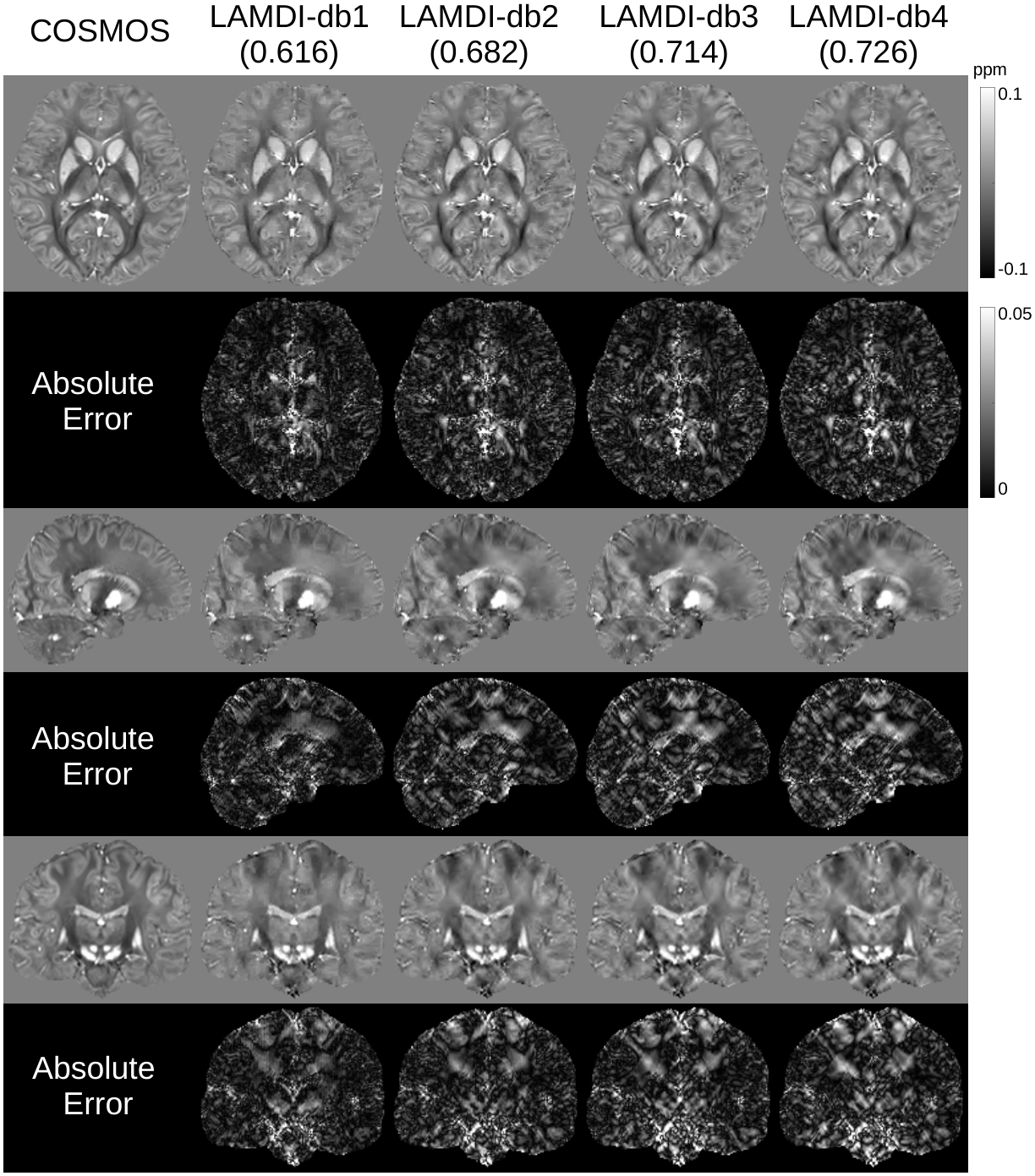}
    \caption{Susceptibility maps reconstructed using COSMOS and LAMDI with different Daubechies wavelets (db1--db4). Among the tested wavelets, db1 produced the highest-quality reconstruction, with the lowest NRMSE (61.6$\%$). As the wavelet order and complexity increase, more high-frequency information is retained in the wavelet coefficients, leading to more visible artifacts and higher NRMSE values (68.2$\%$, 71.4$\%$, and 72.6$\%$ for db2--db4, respectively).}
    \label{fig:compare_approaches_db}
\end{figure}

Post-hoc paired Wilcoxon signed-rank tests were also performed between the best-performing method and each competing method on the held-out subjects S6--S8. The results are given in Table S2 in the Supporting Information. For NRMSE, FANSI tuning achieved the lowest mean value, all three comparisons yielded $p=0.2500$. For HFEN, MEDI tuning achieved the lowest mean value, all three comparisons yielded $p=0.2500$. For SSIM, AMP-PE-L1 achieved the highest mean value, the corresponding p-values were $0.5000$, $0.2500$, and $0.2500$. No post-hoc comparison reached statistical significance.

This comparison highlights the distinction between reconstruction performance and practical usability. FANSI and MEDI benefited from access to COSMOS-based reference susceptibility maps during tuning, allowing their regularization strength to be explicitly adapted to the dataset. In many practical QSM settings, however, such reference maps are unavailable, and repeated parameter retuning across datasets, scanners, or preprocessing pipelines may be cumbersome or infeasible. In contrast, AMP-PE-L1 and LAMDI avoid this requirement by estimating their parameters directly from the observed data.

\subsection{The choice of wavelet bases}

There are many wavelets in the Daubechies family, and db1$\sim$db10 are commonly used in image processing \cite{DBWav92}. As the wavelet order and complexity increase, more high-frequency information tends to be retained in the wavelet coefficients $\vv$. Because streaking artifacts in QSM are dominated by high-frequency components, we chose the simplest member of this family, the db1 (Haar) wavelet, to better suppress such high-frequency artifacts. Figure~\ref{fig:compare_approaches_db} compares the susceptibility maps of Subject 6 obtained by LAMDI using different wavelets. Among them, the db1 wavelet produced the susceptibility map with the fewest visible artifacts and the lowest NRMSE. In addition, when the high-pass db1 filter $\vh_{\textnormal{high}}=[1\ -1]$ is applied to an image, the resulting coefficients correspond essentially to differences between adjacent voxels, i.e., image gradients. In this sense, minimizing the $l_1$-norm of the db1 wavelet coefficients is closely related to TV minimization.

\section{Discussion}

The relatively lower performance of FANSI and MEDI in NRMSE and SSIM may be partially related to their parameter-selection strategies. In our experiments, FANSI used the L-curve method for parameter selection, whereas MEDI used the default parameter recommended by the toolbox. These strategies do not explicitly estimate the regularization strength from the observed data in the same way as AMP-PE-based methods, which may lead to suboptimal trade-offs between artifact suppression and structural preservation. In addition, MEDI depends on a morphology mask derived from the magnitude image. Its performance is thus influenced by the quality of that mask and by the degree to which magnitude boundaries match the true susceptibility boundaries. FANSI, on the other hand, applies TV regularization more uniformly and may therefore oversmooth regions with intrinsically large susceptibility gradients.

This highlights a distinction of AMP-PE-L1 and LAMDI: both methods estimate their regularization parameters automatically from the observed data and therefore avoid the need for heuristic parameter selection or manual tuning. As a result, they may reduce sensitivity to dataset-specific regularization parameters and may be more convenient when applying the same reconstruction framework across subjects and acquisition settings. This practical advantage is useful in QSM, where the optimal regularization strength may vary with image quality, preprocessing errors, and acquisition protocol.

Although AMP-PE-L1 achieved the best average NRMSE and SSIM, its HFEN performance was substantially worse than that of the other methods. This behavior is consistent with the limitation discussed earlier for the single-Laplace prior. Because the empirical distribution of QSM wavelet coefficients is more heavy-tailed than a single Laplace distribution, AMP-PE-L1 tends to overestimate the regularization parameter, leading to blocky or pixelated reconstructions. Such effects may have a relatively modest impact on global similarity measures such as NRMSE and SSIM, but they are penalized much more strongly by HFEN due to the loss of fine structural detail.

Compared with AMP-PE-L1, the proposed LAMDI uses a more flexible two-component Laplace-mixture prior that better matches the heavy-tailed coefficient distribution. As a result, LAMDI achieved similar SSIM and moderately higher NRMSE than AMP-PE-L1, while substantially reducing HFEN. This result is consistent with our hypothesis that a more flexible prior model can alleviate the limitation of AMP-PE-L1, namely, its tendency to oversmooth or pixelate the susceptibility map due to prior mismatch. In this sense, LAMDI offers a better balance between global reconstruction fidelity and local detail preservation, while retaining its main advantage in automatic parameter estimation.

The original AMP-PE-L1 paper recommended the db1 or db2 wavelet bases for QSM. Table S3 in the Supporting Information compares the NRMSE, HFEN, and SSIM of the susceptibility maps reconstructed by AMP-PE-L1 using these two wavelet bases, and Figure S1 in the Supporting Information presents a visual comparison for Subject 6. The db1 wavelet yielded lower NRMSE and higher SSIM, whereas the db2 wavelet produced a lower HFEN. A plausible explanation is that the db2 wavelet retains additional high-frequency content, which can partially compensate for the loss of fine detail caused by over-regularization. However, this additional high-frequency content also includes residual streaking artifacts. As a result, the db2 wavelet leads to a lower HFEN, despite producing less favorable NRMSE and SSIM than db1. Notably, even with the db2 wavelet basis, the HFEN of AMP-PE-L1 remained higher than that achieved by the proposed LAMDI.

\begin{table}[tbp]
\caption{The estimated parameters and corresponding data log-likelihoods obtained from Subject 6 using AMP-PE-L1 and LAMDI.}
\vspace{1em}
\label{tab:parameter}
\centering
\begin{tabular}{lccccccccc}
\toprule
 &\multicolumn{4}{c}{Signal Prior} & & \\ \cmidrule(lr){2-5}
\multirow{-2}{*}{Method} & $\eta_1$ & $\lambda_1$ & $\eta_2$ & $\lambda_2$ & \multirow{-2}{*}{Noise Prior $\tau$}  &\multirow{-2}{*}{Log-likelihood} \\ \midrule
AMP-PE-L1 & 1 & $94.7$ &-- & -- &$661.5$  &$-8.91\times 10^6$  \\
LAMDI & $0.74$ & $7.96\times 10^4$ & $0.26$ & $31.4$ & $735.4$ & $-8.65\times 10^6$ \\
\bottomrule
\end{tabular}
\end{table}

To quantitatively assess whether the proposed two-component Laplace-mixture prior provides a better model for the wavelet coefficient distribution, we compared the estimated parameters and corresponding data log-likelihoods obtained from Subject 6 using AMP-PE-L1 and LAMDI (Table~\ref{tab:parameter}). LAMDI achieved a higher likelihood than AMP-PE-L1, indicating that the Laplace-mixture prior provides a better fit to the empirical distribution of the wavelet coefficients $\vv$. This finding is consistent with our central hypothesis that the coefficient distribution is heavy-tailed and may not be fully characterized by the single-Laplace prior used in AMP-PE-L1, which models all coefficients with a single decay parameter, $\lambda=94.7$.

By contrast, LAMDI decomposes the coefficient distribution into two distinct components. Approximately $\eta_2=26\%$ of the coefficients are assigned to the second Laplace component, which captures the long-tail portion of the distribution with a relatively small decay parameter,  $\lambda_2=31.4$. These coefficients correspond to sharp susceptibility transitions or other high-contrast structures. The remaining $\eta_1=74\%$ of the coefficients are modeled by the first component, which has a much larger decay parameter, $\lambda_1=7.96 \times 10^{4}$, indicating that it primarily captures the large number of near-zero coefficients in relatively homogeneous regions. In this way, the two components separate small and large coefficients, allowing LAMDI to model both the sparse bulk of the distribution and its heavy tail more flexibly.

\section{Conclusion}

In this paper, we proposed an automatic regularization-parameter-estimation approach for QSM dipole inversion, termed Laplace-Mixture Dipole Inversion (LAMDI). By replacing the single-Laplace prior in AMP-PE with a two-component Laplace mixture prior, the proposed method provides a more flexible model for the heavy-tailed distribution of image gradients in susceptibility maps. The model parameters were estimated automatically within the AMP-PE framework using an Expectation–Maximization procedure. Experimental results on one public multi-orientation QSM dataset showed that LAMDI achieves competitive reconstruction performance while retaining the practical advantage of automatic parameter estimation. Compared with AMP-PE-L1, LAMDI produced comparable NRMSE and SSIM, but substantially reduced HFEN and visible block-like artifacts. Compared with FANSI and MEDI under standard default parameter-selection settings, LAMDI achieved favorable HFEN and competitive NRMSE/SSIM without manual tuning.

Overall, the proposed LAMDI approach provides an effective automatic parameter-estimation alternative for QSM dipole inversion. In future work, we will extend the method to more advanced multi-echo formulations, investigate richer prior models, and evaluate its performance on broader datasets and clinical applications.

\newpage

\begin{appendices}

\section{Linearization of the measurement model}
\label{app:sec:linearization}
As discussed in \cite{NonlinearMEDI:Liu:2013}, letting $\boldsymbol\chi^{(l)}$ denote the susceptibility in the $l$-th iteration, a linear approximation of the measurement $\vy_e$ in the next $(l+1)$ iteration can be obtained using the Taylor series:
\begin{align}
\label{eq:exp_measurement_model_linearized}
    \mW_e\cdot\exp(i\boldsymbol\phi_e) = i\mW_e\cdot\exp\left(i\mA_e\boldsymbol\chi^{(l)}\right)\cdot\mA_e\boldsymbol\chi - g\left(\boldsymbol\chi^{(l)}\right) + \vu\,,
\end{align}
where $g\left(\boldsymbol\chi^{(l)}\right)=\mW_e\cdot\exp\left(i\boldsymbol\mA_e\boldsymbol\chi^{(l)}\right)\cdot\left(i\mA_e\boldsymbol\chi^{(l)}-\boldsymbol 1\right)$ is a relative constant that depends on $\boldsymbol\chi^{(l)}$. Rewriting the above \eqref{eq:exp_measurement_model_linearized} in the form of a linear measurement model, we have

\begin{align}
\label{eq:exp_measurement_model_linearized_rewrite}
    \vy_e^{(l)} &= \mA_e^{(l)}\boldsymbol\chi + \vu\,,
\end{align}
where $\vy_e^{(l)}$ and $\mA_e^{(l)}$ are the resulting linear measurements and operator respectively:
\begin{align}
\vy_e^{(l)} &= \mW_e\cdot\exp(i\boldsymbol\phi_e) + g\left(\boldsymbol\chi^{(l)}\right)\\
\mA_e^{(l)} &= i\mW_e\cdot\exp\left(i\mA_e\boldsymbol\chi^{(l)}\right)\cdot\mA_e.
\end{align}

To mitigate the influence of corrupted measurements arising from phase-unwrapping and background-field–removal errors, we update the weighting matrix $\mW_e$ at each iteration using the voxel-wise residual $\boldsymbol\epsilon^{(l)}$:
\begin{align}
    \boldsymbol\epsilon^{(l)} = \mW_e\left(\exp(i\boldsymbol\phi_e)-\exp\left(i\mA_e\chi^{(l)}\right)\right).
\end{align}
Let $\sigma^{(l)}_\epsilon$ denote the standard deviation of the residual $\boldsymbol\epsilon^{(l)}$. A voxel is deemed corrupted if its residual magnitude exceeds $6\sigma^{(l)}_\epsilon$. For such voxels, the corresponding data weight is down-weighted and recomputed as
\begin{align}
W_e(n) = \frac{M_e(n)}{\left(\left|\epsilon^{(l)}(n)\right|/\sigma^{(l)}_\epsilon\right)^2}\,,
\end{align}
where $M_e(n)$ and $\epsilon^{(l)}(n)$ are the magnitude (image) and residual at voxel $n$.

\section{Approximate message passing with parameter estimation}
\label{app:sec:amp_pe}
Approximate message passing (AMP) was originally developed as a probabilistic approach for solving linear inverse problems under the assumption that the relevant distribution parameters are known. In our prior work \cite{PE_GAMP17}, we introduced the AMP with parameter estimation (AMP-PE) framework, in which the distribution parameters $\eta_s,\lambda_s$ and $\tau$ are treated as unknown and jointly estimated together with the signal of interest—here, the susceptibility map $\boldsymbol\chi$. A summary of the resulting AMP-PE approach is provided in Algorithm~\ref{alg:amp_pe}.

\begin{algorithm}[tbp]
\caption{The AMP-PE algorithm }\label{alg:amp_pe}
\begin{algorithmic}[1]
\Input $\hat{\vv}^{(0)}, \tau_v^{(0)}, \vs^{(0)};\tau^{(0)},\lambda_s^{(0)}, \eta_s^{(0)}; \alpha, \beta$
\For{$t=\{0,1,\cdots,T\}$}
	\State Output \emph{linear} update: For each $m=1,\cdots,M$
	\begin{subequations}
	\begin{align}
	{\tau_q}^{(t+1)}&=\frac{1}{M}\|\widetilde{\mA}\|_F^2\cdot {\tau_v}^{(t)}\\
	\vq^{(t+1)}&=\widetilde{\mA} \vv^{(t)}-{\tau_q}^{(t+1)}\cdot \vs^{(t)}\,.
	\end{align}
	\end{subequations}
	\State Estimate the undamped output noise distribution parameter $\widehat{\tau}^{(t+1)}$ using \eqref{eq:noise_variance_est}.
    \State Calculate the damped output channel parameter $\tau^{(t+1)}$ with the damping rate $\beta$ using \eqref{eq:noise_variance_est_damping}.
	\State Output \emph{nonlinear} update:
	\begin{subequations}
	\begin{align}
	\label{eq:compute_tau_s_m}
	{\tau_s}^{(t+1)}&=\frac{1}{\tau^{(t+1)}+\tau_q^{(t+1)}}\\
	\label{eq:compute_s_m}
	\vs^{(t+1)}&=\left(\vy-\vq^{(t+1)}\right)\cdot {\tau_s}^{(t+1)}.
	\end{align}
	\end{subequations}
	\State Input \emph{linear} update: For each $n=1,\cdots,N$
	\begin{subequations}
	\begin{align}
	{\tau_r}^{(t+1)}&=\left[\frac{1}{N}\|\widetilde{\mA}\|_F^2\cdot {\tau_s}^{(t+1)}\right]^{-1}\\
	\vr^{(t+1)}&=\vv^{(t)}+{\tau_r}^{(t+1)}\cdot\widetilde{\mA}^T \vs^{(t+1)}\,.
	\end{align}
	\end{subequations}
	\State Estimate the undamped input signal prior parameters $\widehat{\eta}_s^{(t+1)},\ \widehat{\lambda}_s^{(t+1)}$ using \eqref{eq:pe_lap_mix_weights}, \eqref{eq:pe_lap_mix_reg}.
    \State Calculate the damped parameters input channel parameters $\eta_s^{(t+1)},\ \lambda_s^{(t+1)}$ with the damping rate $\beta$ using \eqref{eq:pe_lap_mix_weights_damping}, \eqref{eq:pe_lap_mix_reg_damping}.
	\State Input \emph{nonlinear} update: Estimate the undamped signal $\widehat{\vv}^{(t+1)}$ using \eqref{eq:est_wave_coef}. 
    The undamped variance $\widehat{\tau}_v^{(t+1)}$ is updated as follows
    \begin{align}
        \widehat{\tau}_v^{(t+1)}=\frac{1}{N}\sum_{n=1}^N\tau_r^{(t+1)}\cdot \mathbbm{1}\left(\widehat{v}^{(t+1)}(n)\neq 0\right),
    \end{align}
    where $\mathbbm{1}(\cdot)$ is an indicator function.
    \State Calculate the damped signal $\vv^{(t+1)}$ and variance $\tau_v^{(t+1)}$ with damping rate $\alpha$.
	\If {$\vv^{(t+1)}$ reaches convergence}
		\State $\vv=\vv^{(t+1)}$ and \textbf{break};
	\EndIf
\EndFor
\State\Return The recovered signal $\vv$;
\end{algorithmic}
\end{algorithm}

\subsection{Input channel parameter estimation}
Assuming the distribution parameters $\{\lambda_s\}_{s=1}^2$ are random variables with uniform priors, their posterior distributions are approximated in AMP-PE as follows
\begin{align}
    p\left(\{\eta_s,\lambda_s\}_{s=1}^2|\vy\right)\propto\prod_{n=1}^N\int_{-\infty}^\infty\sum_{s=1}^2\eta_s\cdot\frac{\lambda_s}{2}\exp(-\lambda_s|v_n|)\cdot\mathcal{N}(v_n|r_n,\tau_r)\ dv_n
\end{align}
where $v_n$ is the $n$-th wavelet coefficients of $\vv\in\mathbb{R}^N$. To simplify the computation, we replace the Gaussian approximation $\mathcal{N}(v_n|r_n,\tau_r)$ with a delta function $\delta(v_n-r_n)$. We then have
\begin{align}
    p\left(\{\eta_s,\lambda_s\}_{s=1}^2|\vy\right)\propto\prod_{n=1}^N\sum_{s=1}^2\eta_s\cdot\frac{\lambda_s}{2}\exp(-\lambda_s|r_n|)\,.
\end{align}
Similarly, by introducing a latent component indicator $z\in\{1,2\}$, with $P(Z=s)=\eta_s$, we can write the complete data likelihood as follows:
\begin{align}
    \prod_{n=1}^N\prod_{s=1}^2\left(\eta_s\cdot\frac{\lambda_s}{2}\exp(-\lambda_s|r_n|)\right)^{\mathbbm{1}(z_n=s)}\,,
\end{align}
where $\mathbbm{1}(\cdot)$ is the indicator function. We can then compute the MAP estimations of the parameters using the EM algorithm, which consists of two steps:

\textbf{E-step}. We compute the posterior of each Laplace component for every $r_n$:
\begin{align}
\label{eq:posterior_laplace_r}
h_s(n)=P(z=s|r_n)=\frac{\eta_s\cdot\lambda_s/2\cdot\exp(-\lambda_s|r_n|)}{\sum_k\eta_k\cdot\lambda_k/2\cdot\exp(-\lambda_k|r_n|)}.
\end{align}

\textbf{M-step}. We maximize the expected log-likelihood function with respect to the posterior of the latent variable $z$. We have the following objective function $g(\cdot)$:
\begin{align}
\begin{split}
    &g\left(\{\eta_s,\lambda_s\}_{s=1}^2\right)\\
    =&\sum_{n=1}^N\sum_{s=1}^2 h_s(n)\cdot\log\left[ \eta_s\cdot\frac{\lambda_s}{2}\cdot\exp(-\lambda_s|r_n|)\right].
\end{split}
\end{align}

The undamped parameters that maximize the above objective function are
\begin{align}
    \label{eq:pe_lap_mix_weights}
    \eta_s &= \frac{\sum_{n=1}^Nh_s(n)}{N}\\
    \label{eq:pe_lap_mix_reg}
    \lambda_s &= \frac{\sum_{n=1}^Nh_s(n)}{\sum_{n=1}^Nh_s(n)|r_n|}.
\end{align}
In practice, the damping operation is also performed on the estimated parameters. Let $\eta_s^{(t)},\lambda_s^{(t)}$ denote the damped parameters at AMP iteration $t$, and let $\widehat{\eta}_s^{(t+1)},\widehat{\lambda}_s^{(t+1)}$ denote the undamped estimated produced by the above \eqref{eq:pe_lap_mix_weights},\eqref{eq:pe_lap_mix_reg} at iteration $t+1$. The damped update is given by
\begin{align}
    \label{eq:pe_lap_mix_weights_damping}
    \eta_s^{(t+1)} = \eta_s^{(t)}+\beta\left(\widehat{\eta}_s^{((t+1)}-\eta_s^{(t)}\right)\\
    \label{eq:pe_lap_mix_reg_damping}
    \lambda_s^{(t+1)} = \lambda_s^{(t)}+\beta\left(\widehat{\lambda}_s^{((t+1)}-\lambda_s^{(t)}\right),
\end{align}
where $\beta\in(0,1]$ is the damping rate for the parameter updates. We empirically choose $\beta=0.1$ for the QSM dipole inversion experiments.

\subsection{Output channel parameter estimation}
The undamped noise variance $\tau$ in the output channel can be estimated by
\begin{align}
\label{eq:noise_variance_est}
    \widehat{\tau}^{(t+1)}=\tau_q^{(t+1)} + \frac{1}{M}\|\vy-\vq^{(t+1)}\|_2^2
\end{align}
The damped noise variance at the $(t+1)$-th iteration is given by
\begin{align}
    \label{eq:noise_variance_est_damping}
    \tau^{(t+1)}=\tau^{(t)}+\beta\left(\widehat{\tau}^{(t+1)}-\tau^{(t)}\right).
\end{align}

\end{appendices}

\bibliographystyle{myunsrt}
\bibliography{Bibliography}

\end{document}